\documentclass[twocolumn,aps,prl,preprintnumbers]{revtex4-2}
\usepackage[utf8]{inputenc}

\usepackage{verbatim}
\usepackage[colorlinks=true,linkcolor=black, citecolor=black,urlcolor=black]{hyperref}
\usepackage{array}
\usepackage{amssymb}
\usepackage{latexsym}
\usepackage{tikz}
\usepackage{subcaption}
\usepackage{mathtools}
\usepackage{shuffle}

\newcommand{\Tr}{{\rm Tr }}

\usetikzlibrary{trees}
\usetikzlibrary{calc}
\usetikzlibrary{decorations.pathmorphing}
\usetikzlibrary{decorations.markings}
\usetikzlibrary{intersections}

\tikzset{
photon/.style={decorate, decoration={snake}},
particle/.style={postaction={decorate},
    decoration={markings,mark=at position .5 with {\arrow{>}}}},
antiparticle/.style={postaction={decorate},
    decoration={markings,mark=at position .5 with {\arrow{<}}}},
gluon/.style={decorate, decoration={coil,amplitude=2pt, segment length=4pt},color=purple},
wilson/.style={color=blue, thick},
scalarZ/.style={postaction={decorate},decoration={markings, mark=at position .75 with{\arrow[scale=1]{stealth}}}},
scalarX/.style={postaction={decorate}, dashed, dash pattern = on 4pt off 2pt, dash phase = 2pt, decoration={markings, mark=at position .53 with{\arrow[scale=1]{stealth}}}},
scalarZw/.style={postaction={decorate},decoration={markings, mark=at position .75 with{\arrow[scale=1]{stealth}}}},
scalarXw/.style={postaction={decorate}, dashed, dash pattern = on 4pt off 2pt, dash phase = 2pt, decoration={markings, mark=at position .60 with{\arrow[scale=1]{stealth}}}}
}

\newtheorem{conj}{Conjecture}
\begin{document}

\title{Cosmic Wheels:\\
    From integrability to the Galois coaction}

\author{\"Omer G\"urdo\u gan}
 \email{Omer.Gurdogan@maths.ox.ac.uk}
\affiliation{%
  Mathematical Institute, University of Oxford,\\ Woodstock Road, Oxford, OX2 6GG, United Kingdom
}%

\date{\today}

\begin{abstract}
  We argue that the description of Feynman loop integrals as
  integrable systems is intimately connected with their motivic
  properties and the action of the Cosmic Galois Group. We show how in
  the case of a family of fishnet graphs, coaction relations between
  them follow directly from iterative constructions of Q-functions in
  the Quantum Spectral Curve formalism. Using this observation we
  conjecture a ``differential equation for numbers'' that enter these
  periods.
\end{abstract}

\maketitle

 \def\rowsep{12pt}

\newcolumntype{R}{>{$}r<{$}}
\newcolumntype{L}{>{$}l<{$}}
\newcolumntype{M}{R@{${}$ \begin{minipage}[t]{1cm}$\displaystyle {}$\end{minipage}}L}

\def\deltatable{
\begin{widetext}
  \begin{table*}
    \begin{tabular}{MM}
      m & \delta_m\\
      \hline\\[3pt]
      $0$ & \displaystyle
            -12\, f_{3}\\[\rowsep]
        $1$ & \displaystyle
              -189\, f_{7} + 288\, f_{3,3}  \\[\rowsep]
        $2$ &\displaystyle
            +\frac{2222721}{40}\, f_{11} + 6804\, f_{3,7} + 6804\, f_{7,3} - 18144\, f_{3,3,3} - 2160\, f_{3,3,5} - 2160\, f_{3,5,3} + 1728\, f_{5,3,3}
              \\[\rowsep]
          $3$ &\displaystyle
          \begin{aligned}[t]
            &-\frac{140451186537}{30800}\, f_{15} - \frac{13336326}{5}\, f_{3,11} + 142884\, f_{7,7} - \frac{13336326}{5}\, f_{11,3}\\
            &- 489888\, f_{3,3,7} - 67320\, f_{3,3,9} - 34020\, f_{3,5,7} - 489888\, f_{3,7,3} -  34020\, f_{3,7,5} 
            - 67320\, f_{3,9,3}+ 47628\, f_{5,3,7} \\
            &  + 47628\, f_{5,7,3}- 489888\, f_{7,3,3} - 34020\, f_{7,3,5} - 34020\, f_{7,5,3}- 13752\, f_{9,3,3} + 1866240\, f_{3,3,3,3} \\
            &+  311040\, f_{3,3,3,5} + 311040\, f_{3,3,5,3} + 124416\, f_{3,5,3,3}- 248832\, f_{5,3,3,3} + 124416\, f_{3,3,3,3,3}\\
          \end{aligned}\\
  \end{tabular}
  \caption{First few wheel periods in an $f$-alphabet.}
\label{tab:periods}
  \end{table*}
\end{widetext}}

\def\wtable{
\begin{widetext}
  \begin{table*}
    \begin{tabular}{MM}
            m & W_{0,m}\\
      \hline\\[3pt]
$1$ &\displaystyle -12\,f_{3} \\[\rowsep]
$2$ &\displaystyle -189\,f_{7} + 144\,f_{3, 2^2}\\[\rowsep]
$3$ &\displaystyle  \frac{2222721}{40}\,f_{11} + \frac{342}{5}\,f_{3, 2^4} + 2268\,f_{7, 2^2} - 1944\,f_{3, 5, 3}\\[\rowsep]
          $4$ &\displaystyle
          \begin{aligned}[t]
            &-\frac{140451186537}{30800} \,f_{15} + \frac{682046379}{3455}\,f_{3, 2^{6}} + \frac{10773}{10}\,f_{7, 2^4} - \frac{6668163}{10}\,f_{11, 2^2} + 13392\, f_{3, 3, 9} - 
            51030\,f_{3, 5, 7} \\
            &- 20088\,f_{3, 9, 3}- 30618\,f_{7, 5, 3} - 51840\,f_{3, 3, 3, 2^3} + 31104\,f_{3, 3, 5, 2^2} + 38880\,f_{3, 5, 3, 2^2} + 
            46656\,f_{3, 3, 3, 3, 3}
          \end{aligned}\\
  \end{tabular}
  \caption{First few coefficients in the $\mu$ expansion of $W_0$.}
\label{tab:wzero}
  \end{table*}
\end{widetext}}



\section{Introduction}
One of the lessons from recent developments in perturbative Quantum
Field Theory is that the most efficient way to deal with Feynman loop
integrals is to avoid them where possible. This point of view is
prevailingly supported by the achievements of the cluster bootstrap
programme for scattering amplitudes in maximally-supersymmetric
Yang-Mills theory (MSYM), in which loop integration is traded with
linear constraints on a linear space of functions conjectured to
contain the amplitude
\cite{Goncharov:2010jf,*Dixon:2016nkn,*Dixon:2011pw,*Drummond:2018caf,*Dixon:2016nkn,*Drummond:2014ffa,*Caron-Huot:2019bsq}. These methods rely heavily on the coalgebra
structure exhibited by Feynman integrals as iterated period
integrals. In particular the symbol, the central ingredient of cluster
bootstrap, is essentially an iteration of the Galois coaction on
multiple polylogarithms \cite{brown2011mixed, Goncharov2002GaloisSO}.

A number of powerful conjectures on periods in QFT are expressed in
terms of a motivic Galois coaction once they are lifted to motivic
periods \cite{brown2011decomposition, brown2017feynman, Brown_2012}. The
(motivic) periods in $\phi^4$ theory are expected to be closed under a
``cosmic'' Galois group that acts on these periods, and this was the
guiding principle of studies of these periods through 11 loops
\cite{Panzer:2016snt}. Dimensionally-regulated one- and two-loop
integrals observe this principle order-by-order in the expansion
around an integer dimension, where the coaction is expressed in terms
of the original graph with cut and contracted edges
\cite{Abreu:2019xep,*Abreu:2019eyg,*Abreu:2018nzy,*Abreu:2017mtm,*Abreu:2017enx}.
Furthermore, evidence for the Galois coaction relations have been
observed in six--particle amplitudes in MSYM evaluated on special
kinematical datapoints \cite{Caron-Huot:2019bsq} and in results for
the electron anomalous magnetic moment. And most recently, the
coaction on general one-loop graphs in four dimensions has been
studied in \cite{tapuskovic2020motivic}. All these studies that reveal
fascinating properties of Feynman periods employ their
often-complicated integral representations, and the underlying
geometry determined by the integrands. More often than not, this
complexity is in contrast with the simplicity of the emergent
structures.

\begin{figure}[]
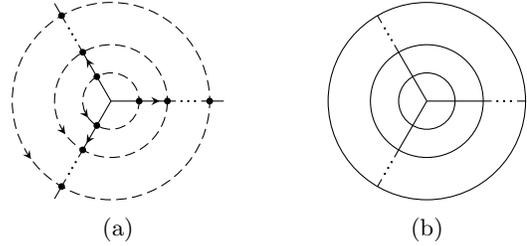

  \centering
  \begin{subfigure}[b]{0.4\linewidth}
  \tikz[scale=0.75]{
    \foreach \rad in {0.5,1,1.75}{
      \draw[scalarXw] (0,0) circle (\rad);
      \foreach \ang in {0,120,240}{
        \draw[black,fill=black] (\ang:\rad) circle (0.05);
        }
    }

    \foreach \t in {0, 120, 240}{
      \draw[scalarZ] (0,0) -- (\t:1.15);
      \draw[black,fill=black] (\t:1.25) circle (0.01);
      \draw[black,fill=black] (\t:1.37) circle (0.01);
      \draw[black,fill=black] (\t:1.49) circle (0.01);
      \draw[] (\t:1.6) -- (\t:2);

       }

     }
     \caption{}
     \label{feynmandiag}
   \end{subfigure}%
   ~ \,\,\,\,\,\,
\begin{subfigure}[b]{0.4\linewidth}
\tikz[scale=0.75]{
    \foreach \rad in {0.5,1,1.75}{
      \draw[] (0,0) circle (\rad);
    }

    \foreach \t in {0, 120, 240}{
      \draw[] (0,0) -- (\t:1.15);
      \draw[black,fill=black] (\t:1.25) circle (0.01);
      \draw[black,fill=black] (\t:1.37) circle (0.01);
      \draw[black,fill=black] (\t:1.49) circle (0.01);
      \draw[] (\t:1.6) -- (\t:1.75);
     }
   }
   \caption{}
      \label{wheelgraph}
   \end{subfigure}
   \caption{(a)A Feynman diagram that contributes to the operator
     (\ref{eq:correlator}) and (b) a corresponding Feynman graph.}
  \label{fig:wheels}
\end{figure}

On the other hand, the description of certain Feynman integrals as
integrable systems paints an entirely different picture
\cite{Gurdogan:2015csr,Zamolodchikov:1980mb,Isaev:2003tk,Loebbert:2020hxk}. This is for example
the case in the Fishnet Model \cite{Gurdogan:2015csr}, which is an
\emph{integrable} QFT of two complex scalars $\phi_1$ and $\phi_2$
interacting with a particular quartic coupling of the form
$\Tr\phi_1\phi_2\phi_1^\dagger\phi_2^\dagger$ . The virtue of this
coupling is that it only allows up to a single Feynman diagram to
contribute to several quantities at each loop order, and this provides
an opportunity to realise such descriptions and the possibility of
computing the periods with no mention of loop integration
whatsoever. The state-of-the-art approach on integrability in QFT is
formulated in terms of the Quantum Spectral Curve
\cite{Gromov:2015vua,Gromov:2013pga} and it has been successfully
adapted to describe observables in the Fishnet Model
\cite{Gromov:2017cja}.
   
The purpose of this Letter is to argue that there is a close link
between the integrability of the Fishnet Model and the coaction
properties of its periods: We will present a problem in which
\emph{coaction relations can be straightforwardly derived from the
  properties of the quantum spectral curve}.

To this end, we will focus on the periods of wheel graphs with three
spokes in the planar limit of the Fishnet Model. These graphs enter
the perturbative expansion of the correlator
\begin{equation}
  \label{eq:correlator}
  \langle \Tr \phi_1^3(0) \,\Tr \bar \phi^3_1(x)\rangle
  =
  \bigl[x^2\bigr]^{-D},\quad D= 3-\xi^3 \delta
\end{equation}
where $D$ is the quantum scaling dimension of the operator
$\Tr \phi_1^3$. The anomalous dimension $\delta$ has a perturbative
expansion in the cube of the fishnet coupling $\xi$:
\begin{equation}
  \label{eq:deltagen}
  \delta
  =
  \sum_{i=0}^{\infty}\, \delta_m \, \xi^{3m}\, .
\end{equation}

This expansion is justified from the diagrams that contribute to these
correlators (depicted in Figure \ref{feynmandiag}) by noting that the
number of interactions is always a multiple of three. In the Feynman
diagram, the solid and dashed lines represent propagators of complex
scalars $\phi_1$ and $\phi_2$, respectively. One of the operators in
(\ref{eq:correlator}) is placed at the origin while the other is
imagined to reside at infinity. Up to a normalisation the numbers
$\delta_m$ are identified \footnote{cfr equation (2.3.9) of
  \cite{Panzer:2015ida} where it is apparent that the definition of
  periods is identical to that of anomalous dimensions} with the
periods of the corresponding ``wheel'' Feynman graph in Figure
\ref{wheelgraph}.

Direct calculations are only available for $\delta_0$
\cite{Broadhurst:1985vq} and for $\delta_1$ \cite{Panzer:2016snt}, and
in general it is proven that they are linear combinations of Multiple
Zeta Values (MZVs) \cite{Brown_2012, Panzer:2015ida}. Owing to the
integrability of the Fishnet Model, powerful techniques of
integrability can be employed to algorithmically determine further
anomalous dimensions $\delta_m$ \cite{Gromov:2017cja} in terms of
these numbers.
  
\section{Solving the $L=3$ Baxter equation recursively}
As detailed in \cite{Gromov:2017cja}, in the QSC setup, the wheel
periods are determined through a quantisation condition 
\begin{equation}
  \label{eq:quant}
 q_2(0,\mu)  q_4(0,-\mu) +  q_2(0,-\mu)  q_4(0,\mu)
  =
  0\, ,
\end{equation}
where the Q-functions $q_2(u,\mu)$ and $q_4(u,\mu)$ satisfy the Baxter equation,
\begin{equation}
  \label{eq:baxter}
  \Box q_\alpha
  = \frac{\mu}{u^3} q_\alpha +\frac{(D-3) (D-1)}{4u^2}q_\alpha\, \quad \alpha=2,4
\end{equation}
and are distinguished by their asymptotics:
\begin{align}
  \label{eq:asympt}
  q_2(u,\mu) &= u^{+D/2-1/2} + \mathcal{O}(u^{-1})\nonumber \\
  q_4 (u,\mu)&= u^{-D/2+3/2} + \mathcal{O}(u^{-1})\,.
\end{align}
The $\Box$ denotes the second-order difference operator:
$\Box f(u) = f(u+i) -2 f(u)+f(u-i)$. The parameter $\mu$ is related to
the fishnet coupling as $\mu^2= \xi^3$ \footnote{In
  contrast with \cite{Gromov:2017cja}, we use $\mu$ to denote the
  parameter $m$ therein.}.

To compute the anomalous dimensions, we consider the expansions of $q_\alpha$
in $\delta$ and $\mu$ in with coefficients $q_\alpha^{i,j}(u)$:
\begin{equation}
  \label{eq:qexpansion}
  q_\alpha(u,\mu)
  =
  \sum_{i=-1}^{\infty}
  \sum_{j=-1}^{\lfloor i/2\rfloor}
  q_\alpha^{i,j}(u)\,
  \mu^i
  \delta^j
  \,,
\end{equation}
with
\begin{align}
  q_2^{i,j}&=0\quad \text{for}\quad i<-1\,\, \text{or}\,\,  j<-1,\nonumber\\
  q_4^{i,j}&=0\quad \text{for}\quad i<0\,\, \text{or}\,\,  j<0,
             \label{eq:ijlimit}
\end{align}
while keeping in mind that $\delta$ also has a $\mu$ expansion as in equation
(\ref{eq:deltagen}). For a similar perturbative analysis of the
Konishi anomalous dimensions in MSYM in the context of QSC see
\cite{Leurent:2013mr}.

The Baxter equation (\ref{eq:baxter}) implies the following equation
for the components of $q_2$ and $q_4$:
\begin{equation}
  \label{eq:baxtercomponent}
  \Box q_\alpha^{i,j} =
  \frac{1}{2 u^2}q_\alpha^{i-2,j-1}
  -  \frac{1}{4 u^2}q_\alpha^{i-4,j-2}
  +   \frac{1}{2 u^3}q_\alpha^{i-1,j}\,,
\end{equation}
which can be algorithmically solved as a linear combination
\footnote{with coefficients at most linear in $u$} of Multiple Hurwitz
Zeta functions (MHZFs):
\begin{equation}
  \eta_{s_1,\dotsc s_k}(u)
  =\sum_{n_1>\dotsm n_k\geq 0}\frac{1}{(u+in_1)^{s_1} \dotsm (u+in_k)^{s_k}}
\end{equation}
using well-known relations that they satisfy
\cite{Gromov:2017cja,Gromov:2015vua,Gromov:2013pga}. Importantly, when
evaluated at $u=\sqrt{-1}$, they reduce to Multiple Zeta Values (MZVs):
\begin{equation}
  \eta_{s_1,\dotsc s_k}(\sqrt{-1})
  =
  (-1)^{\frac{1}{2}\sum_{i=1}^ks_k}\zeta_{s_1,\dotsc s_k}\,.
\end{equation}

\deltatable

\section{Patterns in nested wheels}
\subsection{General structure}
The poles in $u$ in the RHS of equation (\ref{eq:baxtercomponent})
imply that $\delta_m$ are linear combinations of MZVs of weight up to
\mbox{$4m+3$} and indices $\{1,2,3\}$ \footnote{It is straightforward
  to write $\eta_w(u)$ in terms of other MHZFs with argument $u+i$,
  and this does not change the alphabet of their indices}.  The first
few numbers, computed in an ``$f$-alphabet''
\cite{brown2011decomposition} using the Maple package
\texttt{HyperlogProcedures} \cite{hyperlogprocedures} are presented in
Table \ref{tab:periods}. The expressions for $\delta_{\leq 3}$ match
the those provided in \cite{Gromov:2017cja} in terms of MZVs, and we
obtained all periods through $\delta_{\leq 7}$. Their lengthy
expressions can be found in the ancillary file \texttt{wheels.m}.

The main point of rewriting the MZVs in an $f$-alphabet is that it is
a basis where the Galois coaction is simply deconcatenation
\cite{brown2011decomposition}:
\begin{equation}
  \label{eq:deconcat}
  \Delta f_w
  =
  \sum_{w_1 w_2 = w}\,f_{w_1} \otimes f_{w_2}\, ,
\end{equation}
provided that $w$ only contains odd letters, while multiplication of such words is the shuffle product
\begin{equation}
  f_{w_1}  f_{w_2}
    =
    \sum_{w \in w_1 \shuffle w _2}f_{w}\,.
  \end{equation}
  Multiplication of a word by $\zeta_2=f_2$ works by appending the
  letter $2$ to it:
 \begin{equation}
   f_2 f_w = f_{w2}\,.
 \end{equation}
 One can also define derivatives that act on the first entry of a word
 in an $f$-alphabet in the following way:
\begin{equation}
  \partial_{2n+1} f_{aw}
  =
    \begin{cases} 
      f_w & a = 2n+1\\
      0 &  a\neq 2n+1\,.
   \end{cases}
 \end{equation}
 Consquently this framework allows one to consider ``differential
 equations'' for MZVs, which we will use to relate various numbers that
 appear in wheel periods.
 
 We first note some glaring properties of these numbers: The depths of
 MZVs that appear in $\delta_m$ increase with $m$, curiously skipping
 multiples of $4$ as shown in Table \ref{tab:depths}.

 \begin{table}[h]
 \begin{tabular}{llllllllll}
   $m$&0&1&2&3&4&5&6&7&$\dotsm$\\
   Maximum depth in $\delta_m$\,\,\,\,&1&2&3&5&6&7&9&10&$\dotsm$
 \end{tabular}
 \caption{Maximum depth of MZVs in f-basis that appear in each $\delta_m$}
 \label{tab:depths}
 \end{table}
 
 Moreover, they include only odd letters and are symmetric in the last
 two letters. While the former is a well-understood consequence of the
 coaction principle and the fact that $\pi$ and $\zeta_2$ not being
 periods of this theory \cite{brown2017feynman, Panzer:2016snt}, the
 latter, which is a property shared by other wheel families in the
 Fishnet Model \footnote{we defer their study to a future work} but
 not generic $\phi^4$ periods \cite{Panzer:2016snt}, still begs for a
 good explanation.
 
 We should emphasise that ``the $f$-alphabet'' satisfying
 (\ref{eq:deconcat}) is not unique, and it is not guaranteed that
 powers of $\pi$ will not appear after a basis change. Nevertheless we
 observe that this is not the case in the bases available in
 \cite{hyperlogprocedures}.  

 \subsection{Wheel MZVs and their coaction}
 The quantisation condition (\ref{eq:quant}) leads to a natural
 decomposition of a given $\delta_m$ as a polynomial in $\delta_{m'<m}$
 with MZV coefficients. To see this we first observe that using
 \begin{equation}
   q_2^{i,-1}q_4^{j+1,0} =    q_2^{j,-1}q_4^{i+1,0}\qquad  \forall i,j
 \end{equation}
 the condition (\ref{eq:quant}) can be rewritten in the following
 form:
\begin{equation}
  \label{eq:orthogonality}
  W(\delta):=\sum_{i=0}^\infty W_i(\mu)\, \delta^i=0\,.
\end{equation}
The coefficients $W_i(\mu)$ are generating functions

\begin{equation}
  W_i(\mu)
 : =
 \sum_{n=0}^\infty W_{i,n}\, \mu^{2n
 }
\end{equation}
of ``wheel MZVs'' $W_{i,n}$ that are defined as bilinear combinations
of $q_2^{i,j}$ and $q_4^{i,j}$, evaluated at $u=0$:
\begin{equation}
W_{m,n}
 =
 4\,
 \sum_{i,j}
 (-1)^{i-\frac{1}{2}} \,q_4^{i,j}(0)\,q_2^{2m+2-i,n-j}(0)\,,
\end{equation}
where the sum is over all pairs ${i,j}$ that give a non-zero
contribution according to equation (\ref{eq:ijlimit}).

It is now easy to see the structure of the polynomials one obtains by
expanding (\ref{eq:orthogonality}) in $\mu$:
\begin{equation}
  \begin{aligned}[b]
W= &W_{0, 0} + \mu^2 (W_{0, 1} + W_{1, 1} \,\delta_{0})\\
   &+  \mu^4 (W_{0, 2} + W_{1, 2} \,\delta_{0} + W_{2, 2} \,\delta_{0}^2 +   W_{1, 1} \,\delta_{1})\\
   &+\mu^6 \bigl(\begin{aligned}[t]
     &W_{0, 3} + W_{1, 3} \,\delta_{0} + W_{2, 3} \,\delta_{0}^2 + W_{3, 3} \,\delta_{0}^3\\
     &+ W_{1, 2} \,\delta_{1} +    2 W_{2, 2} \,\delta_{0} \,\delta_{1} + W_{1, 1} \,\delta_{2}\bigr)
     \end{aligned}\\
&+ \mathcal{O}(\mu^8)\,.
\label{eq:polyexplicit}
   \end{aligned}
    \end{equation}
    Requiring the vanishing of (\ref{eq:polyexplicit}) at each order,
    and noting that $W_{1,1}=-1$ allows us to express any $\delta_m$
    as a polynomial in $\delta_{m'<m}$ with MZV coefficients.

    As a first iterative pattern, we note that the only new numbers
    that enter a period $\delta_m$ are the coefficients of the powers
    of $\delta_0$, while coefficients of other monomials will have
    occured as the coefficient of some power of $\delta_0$ in a
    smaller wheel.

We will now show that the $W_i$ have curious properties under the
coaction that they inherit from the solutions of the Baxter
equation. As an explicit illustration of these, we will focus on the
sequence $W_{0,m}$. First few elements of this sequence are presented
in Table \ref{tab:wzero}.

\begin{conj}
  The components $q_2^{i,-1}= q_4^{i,0}$ of solutions to the system of equations (\ref{eq:baxter}) \& (\ref{eq:asympt}) satisfy
  \begin{equation}
    \label{eq:derivativeofq}
    c_k^{-1}\partial_{k} q_4^{i,0}(0)
    =
    \begin{cases}
      c_{k-4}^{-1} \,\partial_{k-4} q_4^{i-2,0}(0) & k = 3 \mod 4 \\
      0 & k =1 \mod 4
      \end{cases}
\end{equation}
with the same rational constant $c_k$ for all $i\geq 2$, with the
first few numbers $c_k$:
\begin{align}
      &(c_3,c_7,c_{11},\dotsc) =\nonumber\\
    &\biggl(1,\frac{63}{4},-\frac{740907}{160},\frac{46817062179}{123200},\nonumber\\
  &\qquad \qquad\frac{2748660980272441072644717}{94776262380800},\dotsc\biggr)\, .
\end{align}
\label{conj:qrelconj}
  \end{conj}

The relation (\ref{eq:derivativeofq}) together with the fact that
$q_2^{i,-1}=q_4^{i,0}$ immediately lead to the differential equations
that hold for any $k = 3 \mod 4$:
\begin{align}
  \label{eq:leadingCAP}
            \mathcal{O}_kW_0 := \bigl(c_k^{-1}\,\partial_{k}  - c_{k-4}^{-1}\,\partial_{k-4}\bigr)W_0&=0
\end{align}
while any other derivative $\partial_{k}$ with $k\neq 3 \mod 4$ annihilates $W_0$.

In other words, the coaction relations between wheel MZVs
(\ref{eq:leadingCAP}) rely on a conjectural identity for the
Q-functions, which we have verified for $m\leq 7$, ie up to weight 31.

  \subsection{A differential equation for wheel MZVs}
  Equation (\ref{eq:leadingCAP}) in fact follows from a remarkable
  relation satisfied by $W_0$. We first note that the differential
  equation (\ref{eq:leadingCAP}) implies that $W_0$ has the form
  \begin{equation}
    W_0 = Z \cdot S,
  \end{equation}
  where $\cdot$ denotes concatenation that is distributive over
  addition, $Z$ has a Taylor expansion in $\mu$ that involves only
  \emph{single} zeta values:
  \begin{equation}
    \label{eq:zetadef}
    Z(\mu)
    =
    \sum_{m=1}^{\infty} c_{4m-1}\, f_{4m-1}\mu^{m},
  \end{equation}
  and $S$ is a function that can be worked out a posteriori. Upon defining
  \begin{equation}
    \label{eq:zetatildedef}
    \tilde Z(\mu)
        =
    \sum_{m=1}^{\infty} \,\tilde c_{4m+1}\, f_{4m+1}\mu^{m} 
  \end{equation}
  with
  \begin{align}
      &(\tilde c_5, \tilde c_9, \tilde c_{13},\dotsc) =\nonumber\\
    &\biggl(1,\frac{31}{3},-\frac{718191}{200},\frac{7408590567067}{484000},\nonumber\\
  &\,-\frac{1276765401753338789096939955193067}{2169303723306720000},\dotsc\biggr)\, ,
\end{align}
and solving for the function $S$, we observe that $W_0 $ obeys (up to
the available order) an almost homogeneous differential equation of
the form:
  \begin{equation}
    \label{eq:didedeltabar}
    W_0
    -
    \frac{1}{\mu^4} Z\cdot
    \bigl(
    Z \cdot \partial_3 \partial_3 + \tilde Z \cdot \partial_5 \partial_3
    \bigr)W_0
    =
   \, Z\cdot P
    \, ,
  \end{equation}
  where $P=P(\mu)$ is a function whose expansion only contains powers of $\pi$:
  \begin{equation}
    P(\mu)
    =
    12
    +144\,\mu^2 \zeta_4+\frac{342}{5}\,\mu^8 \zeta_8+\dotsm\,.
  \end{equation}
  Since the powers of $\zeta_2$ cancel out in $\delta_m$, we can
  ignore the $\mathcal{O}(\mu)$ terms in $P$.
  
  To emphasise the predicivity of (\ref{eq:didedeltabar}) we remind
  that both $Z$ and $\tilde Z$ are $\mathcal{O}(\mu)$. In summary,
  \emph{the coactions on all $W_{k,0}$ are encoded in a pair of
    functions whose expansion contains only single zeta values}.  This
  is reminiscent of an observation made about string-theory amplitudes
  considered in \cite{Schlotterer:2012ny, *Drummond:2013vz}
  
  While we focussed on the simplest case of $W_0(\mu)$ in this work,
  similar (and more involved) constructions are in principle possible
  for other sequences of wheel MZVs. For example the analogue of
  equation (\ref{eq:leadingCAP}) for $W_1$ is
  \begin{equation}
    \label{eq:subleadingCAP}
    \bigl(\mathcal{O}_k - \mu^2 \mathcal{O}_{k-4}\bigr)W_1=0\,,
  \end{equation}
  and this relation relies on a relation satisfied by ``higher''
  components of the Q-functions:
  \begin{align}
    \bigl(
    \mathcal O_k-2 \mu^2     \mathcal O_{k-4}   +\mu^4  \mathcal O_{k-8}
    \bigr)
    q_4^{i,1}
    &=0\nonumber
    \\
    \bigl(
        \mathcal O_k-2\mu^2     \mathcal O_{k-4}   +\mu^4  \mathcal O_{k-8}
    \bigr)
    q_2^{j,0}
    &=0, 
  \end{align}
  for all $k=3\mod 4$ in all the available data.

  We conclude the discussion with a comment on the prospects of
  generalisations of (\ref{eq:leadingCAP}) and
  (\ref{eq:subleadingCAP}).  One may be tempted to speculate equations
  for $W_i$ using operators $\mathcal{O}^{(i)}_{k}$ that are defined
  recursively as
  \begin{equation}
       \mathcal{O}^{(i)}_{k}:=       \mathcal{O}^{(i-1)}_{k}-      \mu^2  \mathcal{O}^{(i-1)}_{k-4},\quad     \mathcal{O}^{(0)}_k=\mathcal{O}_k\, .
     \end{equation}
     However, given the range of depths that enter these equations
     being comparable to the amount of data available up to weight 31
     does not allow for convincing checks. Therefore, it will be
     cruicial to prove Conjecture \ref{conj:qrelconj}, and its
     generalisations to make further statements on the coaction on
     wheel periods.  \wtable
  \subsection{Subleading-weight pieces as polynomials in $\delta_m$}
  Separating the anomaous dimensions into the leading- and
  subleading-weight parts:
\begin{equation}
  \delta_m
  =
  \bar{\delta}_{m}
  + (-2)^m\,\tilde{\delta}_{m}
\end{equation}
where $\tilde{\delta}_{m}$ contains all terms of weight \emph{less
  than} \mbox{$4m-3$}, gives further insight to patterns that relate
various anomalous dimensions to each other.

We observe that all known $\tilde{\delta}_m$, can be expressed in
terms of $\delta_{m'}$ with $m'<m$:
\begin{equation}
  \begin{aligned}[b]
    \tilde{\delta}_1=&
    -\delta_0^2\\
    \tilde{\delta}_2=&
    -8\,\delta_1\,\delta_0+\frac{1}{2}(5\delta_0^2-4\,\delta_1)\,\delta_0\\
    \tilde{\delta}_3=&
    -16\,\delta_{2}\, \delta_{0} +4\,(7\, \delta_{0}^2  - 4 \delta_{1})\,\delta_{1}-7 \delta_{0}^4\\
    \tilde{\delta}_4=&
    -40\,\delta_{3}\, \delta_{0} + 10\, (9 \delta_{0}^2 - 4\, \delta_{1})\,\delta_{2} \\
    & -30\, (4 \delta_{0}^2- 3 \delta_{1})\, \delta_{1}\, \delta_{0}+21 \,\delta_{0}^5\\
     \tilde{\delta}_5=&
     -96 \,\delta_4\,\delta_{0} + 24\, (11 \,\delta_{0}^2 - 4 \,\delta_{1})\,\delta_3 -88  (5 \,\delta_{0}^2 - 6 \,\delta_{1})\,\delta_{0}\,\delta_2\\
         &\quad-  44 (15 \,\delta_{0}^2 - 2 \,\delta_{1}) \,\delta_{1}^2 -  48 \,\delta_{2}^2-33(2 \,\delta_{0}^2- 15 \,\,\delta_1)\,\delta_0^4\\
    \,\dotsc
\end{aligned}
\end{equation}
This motivates the following \emph{all-loop} conjecture:

\begin{conj}
  \label{conj:subleading}
  The subleading-weight parts of the anomalous dimensions
  $\tilde\delta_m$ are always polynomials in $\tilde\delta_{m'< m}$, and are
  given by the following formula:
  \begin{multline}
  \tilde\delta_m
  =
  \frac{(2m)!}{(m+1)!} \frac{(\delta_0)^m}{m!}\\
  +
    (-2)^{m-A+1}
  \frac{m\,(2m-1)!}
  {(2m-A+1)!}\,
  \sum_{\mathbf{a}:\, A>|\mathbf{a}|}
  \delta^{\mathbf{a}}
\end{multline}

The sum is over all multiplets
$\mathbf{a}=(a_1,\dotsc,a_{|\mathbf{a}|})$ of positive integers
subject to the condition
\mbox{$ A=\sum_{i=1}^{|\mathbf{a}|}a_i>|\mathbf{a}|$} and we used the
shorthand using words $\mathbf{a}=(a_1,a_2,\dotsc)$:
\begin{equation}
  \delta^{\mathbf{a}}
  =
  \prod_{i=1}^{|\mathbf{a}|}
  \frac{(\delta_{i-1})^{a_i}}{a_i!}\,,
\end{equation}
where $|\mathbf{a}|$ denotes the length of $\mathbf{a}$, and
$A = \sum_{i=1}^{|\mathbf{a}|}{a_i}$ the sum of its letters, ie the
degree of $\delta^{\mathbf{a}}$. 
\end{conj}
We verified Conjecture \ref{conj:subleading} through $\tilde\delta_7$,
ie up to weight 30.
\section{Discussion}
In this Letter we demonstrated evidence for that the integrability of
fishnet theory has a direct consequence on the properties of its
periods under the Galois coaction. We identified a coaction principle
that controls a sequence of numbers that enter the anomalous
dimensions of the operator $\Tr \phi_1^3$ and therefore a family of
nested wheel diagrams. Using these observations we constructed a
powerful equation satisfied by the these numbers, showing that
complicated MZV expressions are in fact encoded in sequence of single
Riemann zeta values. Furthermore we identified other recursive
features of these numbers: We presented a structure that shows that
numbers associated with smaller wheels recur naturally in the results
for larger ones. We also made a conjecture determining the parts of
anomalous dimensions of subleading transcendentailty weight as
polynomials in those of lower loop order.

While we analysed the anomalous dimenions order by order in the
perturbative expansion of the anomalous dimension $\delta$, it would
be very much in the spirit of integrability to make non-perturbative
statements about its coaction. In particular the coaction on the
resummed quantity $W_0(\mu)$ is reminiscent of that of hypergeometric
functions, and we also remark that a non-physical toy model of
(\ref{eq:baxter}) in \cite{Gromov:2017cja} is solved in terms of these
functions, whose coaction has been recently considered in
\cite{Abreu:2019xep, brown2020lauricella}. The relationship between
the difference equations that these functions satisfy, and their
Galois coaction is not yet clear.

We would like to point out that the cancellation of the terms
involving powers of $\pi$ in $W_i$ in equation
(\ref{eq:orthogonality}) relies on conspiracies between the periods
$\delta_m$ and on combinatorial identities for MZVs in $\{1,2,3\}$
. These would be generalisations of identities considered in number
theory literature \cite{1998math.....12020B,Charlton_2015}, and their
proofs would be a key to advancing the observations presented here.

An obvious question, and one of the main motivations of this letter,
is on the possibility of reversing the logic presented: The Galois
coaction on periods is believed to be a universal property of QFTs,
and coactions on Feynman amplitudes are expected to satisfy relations
that may be expressed as differential equations. It would be an
exciting prospect to recast such relations into a difference equation
such as (\ref{eq:baxtercomponent}), which could in turn imply
previously-unknown integrable structures in QFTs.

\begin{acknowledgments}
  The author is grateful to Francis Brown for invaluable suggestions
  that led to results presented in this Letter. The author also would
  like to thank to Nikolay Gromov and the authors of
  \cite{Gromov:2017cja} for assistance on the intricacies of the
  perturbative expansion of QSC and sharing their code, Oliver Schnetz
  for making available the bases of MZVs up to weight 31, Erik Panzer
  for numerous stimulating discussions; and Francis Brown, Nikolay
  Gromov and James Drummond for their comments on the earlier versions of this
  manuscript. This project has received funding from the European
  Research Council (ERC) under the European Union’s Horizon 2020
  research and innovation programme (grant agreement No. 724638)
\end{acknowledgments}
\end{document}